\documentclass[floatfix,twocolumn,aps,showpacs]{revtex4}
\usepackage{amssymb}
\usepackage{graphics,epsf}
\newcommand{\be}{\begin{equation}}
\newcommand{\ee}{\end{equation}}
\newcommand{\bq}{\begin{eqnarray}}
\newcommand{\eq}{\end{eqnarray}}
\newcommand{\bs}{\begin{eqnarray*}}
\newcommand{\es}{\end{eqnarray*}}
\newcommand{\bg}{\bar{g}}
\newcommand{\bl}{\bar{l}}
\newcommand{\bn}{\bar{n}}
\newcommand{\bm}{{\bar{m}}}
\newcommand{\bmt}[1]{\mbox{\boldmath $#1$}}
\newcommand{\nn}{\nonumber}
\begin{document}
\title{Physical interpretation of gauge invariant perturbations of spherically symmetric space-times.}
\author{Brien C Nolan\footnote{Electronic address:
brien.nolan@dcu.ie}}\affiliation{School of Mathematical Sciences,
Dublin City University, Glasnevin, Dublin 9, Ireland.}
%\date{\today}
\begin{abstract}
By calculating the Newman-Penrose Weyl tensor components of a
perturbed spherically symmetric space-time with respect to
invariantly defined classes of null tetrads, we give a physical
interpretation, in terms of gravitational radiation, of odd parity
gauge invariant metric perturbations. We point out how these gauge
invariants may be used in setting boundary and/or initial
conditions in perturbation theory.
\end{abstract}
 \maketitle
%_____________________________________________________________________
%_____________________________________________________________________
\newtheorem{assume}{Assumption}
\newtheorem{theorem}{Theorem}
\newtheorem{prop}{Proposition}
\newtheorem{corr}{Corollary}
\newtheorem{lemma}{Lemma}
\newtheorem{definition}{Definition}
\newcommand{\so}{{\cal{O}}}
\newcommand{\ch}{{\cal{H}}}
\newcommand{\cf}{{\cal{F}}}
\newcommand{\cm}{{\cal{M}}}
\newcommand{\pnc}{{\cal{N}}}
%_______________________________________________________________________
%______________________________________________________________________
\section{Introduction}
Perturbation theory in general relativity is complicated by the
issue of co-ordinate freedom in the unperturbed background
space-time $(M,\bg_{\mu\nu})$. If one formally adds a perturbation
to the metric $\bg_{\mu\nu}\to
g_{\mu\nu}=\bg_{\mu\nu}+h_{\mu\nu}$, it is not necessarily true
that one has moved to a different space-time: $g_{\mu\nu}$ may be
the metric $\bg_{\mu\nu}$ written in a different co-ordinate
system, which is related to the original co-ordinates by an
infinitesmal co-ordinate transformation. This is known as the
identification gauge problem. The gauge freedom represented by
such infinitesmal co-ordinate transformations must be dealt with
carefully. One way to do this is to treat the perturbation problem
in hand using identification gauge invariant (i.g.i.) quantities
\cite{stewart-walker}. For perturbations of spherically symmetric
space-times, a complete set of such quantities representing metric
and matter perturbations, and the corresponding i.g.i.\
perturbation equations, have been given by Gerlach and Sengupta
(GS) \cite{GS}. We review their formalism in Section II below.

This formalism has been applied in many different areas, for
example in studies of non-spherical stellar collapse
\cite{harada,lockitch,jm+carsten1}, critical collapse
\cite{garfinkle+gundlach,frolov,jm+carsten2,gundlach},
phenomenology of naked singularities \cite{hin}, black holes
\cite{sarbach1,allen,siino1}, cosmology
\cite{kodama,tomita1,ishibashi}, nonlinear perturbation theory
\cite{nakamura} and perturbations of gauge fields
\cite{moreno,sarbach2}. These studies have generally extracted the
physical significance of the metric perturbations, e.g.\ by
calculating the radiated power of gravitational waves \cite{hin}
or by making the connection with the more familiar
Regge-Wheeler-Zerilli and Teukolsky perturbation formalisms
\cite{sarbach1}. Nevertheless, a general and direct interpretation
of the full set of i.g.i.\ metric perturbations has not been
given. The aim in the present paper is to attempt to do so by
calculating the Newman-Penrose (NP) Weyl tensor components of the
perturbed space-time. The type-$N$ component, which represents
transverse gravitational waves has previously been calculated in
\cite{hin} and \cite{sarbach1}. In carrying out this calculation,
one encounters another type of gauge problem, namely the freedom
of choice in the null tetrad of the perturbed space-time.

Stewart and Walker \cite{stewart-walker} discussed this additional
gauge invariance, and concluded that the only Weyl scalars that
are both i.g.i.\ and tetrad gauge invariant (t.g.i.) are the
type-$N$ terms, and furthermore, that these terms can only be
gauge invariant if the background is of Petrov type-$D$ or
conformally flat. These include all spherically symmetric
space-times. (We use the phrase `gauge invariant' to refer to a
quantity which is both tetrad and identification gauge invariant.)
Consequently, any attempt to attach physical significance to the
full set of perturbed Weyl scalars seems doomed. However, as we
will see below, this is not the case for odd perturbations (see
\cite{GS} and Section II below). In this case, there is sufficient
geometric information in the background that is invariant with
respect to the generators of odd perturbations to enable the
construction of gauge invariant perturbed Weyl scalars. This will
allow the interpretation of the metric perturbations in terms of
longitudinal and transverse waves propagating in the inward and
outward radial null directions of the spherically symmetric
background and in terms of a perturbation of the Coulombic
interaction. As in the analysis of \cite{stewart-walker}, this
will involve the choice of a special class of tetrads, but one
which admits an i.g.i. description. We follow the curvature,
tetrad and NP conventions of \cite{stewart}.

\section{The Gerlach-Sengupta Formalism}
For convenience, we give a brief review of the formalism
introduced by Gerlach and Sengupta \cite{GS}, following the
presentation of Martin-Garcia and Gundlach \cite{jm+carsten2}. The
metric of a spherically symmetric space-time $M^4$ can be written
as \be ds^2=g_{AB}(x^C)dx^Adx^B+r^2(x^C)\gamma_{ab}dx^adx^b,
\label{bgmetric}\ee where $g_{AB}$ is a Lorentzian metric on a
2-dimensional manifold with boundary $M^2$ and $\gamma_{ab}$ is
the standard metric on the unit 2-sphere $S^2$. Capital Latin
indices represent tensor indices on $M^2$, and lower case Latin
indices are tensor indices on $S^2$. $r(x^C)$ is a scalar field on
$M^2$. 4-dimensional space-time indices will be given in Greek.
The covariant derivatives on $M^4$, $M^2$ and $S^2$ will be
denoted by a semi-colon, a vertical and a colon respectively.
$\epsilon_{AB}$ and $\epsilon_{ab}$ are covariantly constant
anti-symmetric unit tensors with respect to $g_{AB}$ and
$\gamma_{ab}$. We define \bq v_A&=&\frac{r_{|A}}{r}, \label{vdef}\\
V_0&=&-\frac{1}{r^2}+2{v^A}_{|A}+3v^Av_A.\label{v0def}\eq Writing
the stress-energy tensor as \be t_{\mu\nu}dx^\mu dx^\nu=
t_{AB}(x^C)dx^Adx^B+Q(x^C)r^2\gamma_{ab}dx^a dx^b,\ee the Einstein
equations of the spherically symmetric background read \bq
G_{AB}=-2(v_{A|B}+v_Av_B)+V_0g_{AB}&=&8\pi
t_{AB}\label{eeq1}\\\frac12 G^a_a=
-{\cal{R}}+{v^A}_{|A}+v^Av_A&=&8\pi Q,\label{eeq2}\eq where
$G^a_a=\gamma^{ab}G_{ab}$ and ${\cal{R}}$ is the Gaussian
curvature of $M^2$.

Spherical symmetry of the background allows us to expand the
perturbed metric tensor in terms of spherical harmonics. Writing
$Y=Y^m_l$ and suppressing the indices $l,m$ throughout, we have
the following bases for scalar, vector and tensor harmonics
respectively: $\{Y\}$, $\{Y_a:=Y_{:a},S_a:=\epsilon_a^bY_{b}\}$
and
$\{Y\gamma_{ab},Z_{ab}:=Y_{a:b}+\frac{l(l+1)}{2}Y\gamma_{ab},S_{a:b}+S_{b:a}\}$.
These are further classified depending on the transformation
properties under spatial inversion ${\vec x}\to -\vec{x}$: a
spherical harmonic with index $l$ is called even if it transforms
as $(-1)^l$ and is called odd if it transforms as $(-1)^{l+1}$. In
the bases above, $Y,Y_{a}$ and $Z_{ab}$ are even and
$S_a,S_{(a:b)}$ are odd.

The perturbation $\delta g_{\mu\nu}$ of the metric tensor can then
be decomposed as \bq \delta g_{AB}&=&h_{AB}Y,\label{tensor}\\
\delta g_{Ab}&=&h^E_AY_{:b}+h_A^OS_b,\label{vector}\\
\delta
g_{ab}&=&r^2K\gamma_{ab}Y+r^2GZ_{ab}+2hS_{(a:b)}.\label{scalar}
\eq The superscripts $E,O$ stand for even and odd respectively.
Note that $h_{AB}$, $\{h_A^E,h_A^O\}$ and $\{K,G,h\}$ are
respectively a 2-tensor, vectors and scalars on $M^2$. A similar
decomposition of the perturbation of the stress-energy tensor is
made:
\bq \delta t_{AB}&=&\Delta t_{AB}Y,\label{emt-tensor}\\
\delta t_{Ab}&=&\Delta t^E_AY_{:b}+\Delta t_A^OS_b,\label{emt-vector}\\
\delta t_{ab}&=&r^2\Delta t^3 \gamma_{ab}Y+r^2\Delta t^2
Z_{ab}+2\Delta tS_{(a:b)}.\label{emt-scalar} \eq In this case,
$\Delta t_{AB}$, $\{\Delta t_A^E,\Delta t_A^O\}$ and $\{\Delta
t^3, \Delta t^2, \Delta t\}$ are respectively a 2-tensor, vectors
and scalars on $M^2$.

A complete set of identification gauge invariant variables is
produced as follows. An infinitesmal co-ordinate transformation on
the background is generated by a vector field $\vec{\xi}$. Again,
we can decompose into even and odd harmonics and consider
separately the transformations generated by the 1-form fields \bq
\bmt\xi^E
&=&\xi_A(x^C)Ydx^A+\xi^E(x^C)Y_{:a}dx^a,\label{evenxi}\\
\bmt\xi^O&=&\xi^OS_adx^a. \label{oddxi}\eq From the transformed
versions of the metric perturbations, one can construct
combinations which are independent of the coefficients of
$\vec{\xi}$. These combinations are then identification gauge
invariant. Writing \be
p_A=h_A^E-\frac{r^2}{2}G_{|A},\label{pvecdef}\ee a complete set of
i.g.i.\ metric perturbations is given by \bq
k_{AB}&=&h_{AB}-2p_{(A|B)},\label{ktendef}\\
k_A&=&h_A^O-h_{|A}+2hv_A,\label{kvecdef}\\
k&=&K+\frac{l(l+1)}{2}G-2v^Ap_A.\label{kscaldef} \eq Similarly, a
complete set of i.g.i.\ stress-energy tensor perturbations may be
constructed. We will not give these here, but refer the reader to
\cite{GS} or \cite{jm+carsten2}. The full set of i.g.i.\
perturbation equations may also be found in these references; we
will not use these equations in the present paper.

An important point to note is that this formalism is incomplete
for $l=0$ and for $l=1$. For $l=0,1$, $G$ and $h$ are not defined,
being coefficients of zero, and so should be considered to be
zero. The same holds for $h_A^E,h_A^O$ when $l=0$. Thus the gauge
invariants cannot be constructed. However it is convenient to use
the same variables (\ref{ktendef})-(\ref{kscaldef}) for all values
of $l$. For $l=0,1$, these variables are only partially i.g.i.\,
and so gauge-fixing is required. This does not affect the
calculation below.

To conclude this section, we point out the existence of a
preferred gauge in which $h=G=h_A^E=0$. This is the Regge-Wheeler
(RW) gauge. This has the advantage that the bare perturbations of
(\ref{tensor})-(\ref{scalar}) match identically the i.g.i.\
perturbations.
%We will work throughout in this gauge.

\section{Null tetrads and Weyl scalars}
It is convenient to introduce co-ordinates
$x^\mu=(\theta,\phi,u,v)$ on the spherically symmetric background,
with $\mu=1-4$ in the order shown. $u,v$ are null co-ordinates on
$M^2$ which we take to increase into the future. Furthermore, we
specify that $u,v$ are respectively retarded and advanced time
co-ordinates, so that $u$ (respectively $v$) labels the future
(respectively past) null cones of the axis $r=0$. Then the
background line element can be written as
\[ ds^2=-r^2(u,v) d\Omega^2 +2e^{-2f(u,v)}dudv,\]
where the only co-ordinate freedom corresponds to the relabelling
$u\to U(u),v\to V(v)$ of the spherical null cones. We introduce
the null tetrad \bq \bar{m}_\mu
&=&\frac{r}{\sqrt{2}}(\delta^1_\mu+i\sin\theta\delta^2_\mu),\\
\bar{m}^*_\mu
&=&\frac{r}{\sqrt{2}}(\delta^1_\mu-i\sin\theta\delta^2_\mu),\\
\bar{n}_\mu&=&e^{-f}\delta_\mu^4,\\
\bar{l}_\mu&=&e^{-f}\delta_\mu^3,\\
\eq so that
\[\bg_{\mu\nu}=2\bl_{(\mu}\bn_{\nu)}-2\bm_{(\mu}\bm^*_{\nu)}.\]
Here and throughout, the overline indicates a background quantity
and the asterisk represents complex conjugation. With respect to
this tetrad, there is only one non-vanishing Weyl tensor
component; \bq \bar{\Psi}_2 &=&
\frac{1}{6r^2}(2re^{2f}(r_{,uv}+rf_{,uv})-1-2e^{2f}r_{,u}r_{,v})\nn\\
&=&\frac16({\cal{R}}+\frac{1}{r}\Box_2
r-\frac{1}{r^2}(1+\chi)),\eq where $\Box_2$ is the d'Alembertian
of $M_2$ and $\chi=g^{AB}r_{,A}r_{,B}$. Under general Lorentz
transformations of the null tetrad, this term is not invariant.
However, due to spherical symmetry, there is an invariant class of
null tetrads, namely that which takes the two real members of the
tetrad to be the repeated principal null directions of the Weyl
tensor (the ingoing and outgoing radial null directions).
Specifying that we always do this, the only allowed Lorentz
transformations are spin-boosts which involve \be {\bl}^\mu\to
a^2{\bl}^\mu,\quad {\bn}^\mu\to a^2{\bn}^\mu\quad {\bm}^\mu\to
e^{2i\omega}{\bm}^\mu,\label{boosts}\ee where $a,\omega$ are
arbitrary. $\bar{\Psi}_2$ is invariant under these
transformations. Henceforth, a null tetrad
$\{{\bm}^\mu,{\bm^{\!*\,\mu}},{\bn}^\mu,{\bl}^\mu\}$ for the
background will always be taken to lie in this class. Without loss
of generality, we can always take ${\bn}^\mu$ to point in the
radial ingoing null direction and ${\bl}^\nu$ to point in the
radial outgoing null direction.

We write a null tetrad of the perturbed space-time as
$\{{\vec{m}},\vec{m}^*,\vec{n},\vec{l}\}$, with \bq
g_{\mu\nu}&=&\bg_{\mu\nu}+\delta g_{\mu\nu}\nn\\
&=&-2m_{(\mu}m^*_{\nu)}+2l_{(\mu}n_{\nu),}\label{tetrad}\eq where
$l_\mu=\bl_\mu+\delta l_\mu$ and similar for other tetrad members.
The condition (\ref{tetrad}) is an under-determined linear system
for the perturbations $\delta l_\mu$ (etc.) in terms of the metric
perturbations, corresponding to the gauge freedom of Lorentz
transformations. In order that the Weyl scalars calculated below
have an invariant meaning, we must choose the tetrad (or more
correctly, class of tetrads) in an invariant way, as was done
above for the background.

The Weyl scalars are given by \bq
\Psi_0&=&C_{\mu\nu\lambda\sigma}l^\mu m^\nu l^\lambda m^\sigma,\label{p0}\\
\Psi_1&=&C_{\mu\nu\lambda\sigma}l^\mu m^\nu l^\lambda n^\sigma,\label{p1}\\
\Psi_2&=&C_{\mu\nu\lambda\sigma}l^\mu m^\nu n^\lambda {m^*}^\sigma,\label{p2}\\
\Psi_3&=&C_{\mu\nu\lambda\sigma}l^\mu n^\nu {m^*}^\lambda n^\sigma,\label{p3}\\
\Psi_4&=&C_{\mu\nu\lambda\sigma}n^\mu {m^*}^\nu n^\lambda
{m^*}^\sigma.\label{p4}\eq With our choice of background tetrad,
we find that these yield \bq
\delta\Psi_0&=&\delta C_{\mu\nu\lambda\sigma}\bl^\mu \bm^\nu \bl^\lambda \bm^\sigma,\label{dp0}\\
\delta\Psi_1&=&-a\bar{\Psi}_2+\delta C_{\mu\nu\lambda\sigma}\bl^\mu \bm^\nu \bl^\lambda \bn^\sigma,\label{dp1}\\
\delta\Psi_2&=&b\bar{\Psi}_2+\delta C_{\mu\nu\lambda\sigma}\bl^\mu
\bm^\nu \bn^\lambda
{\bm}^{\!*\,\sigma},\label{dp2}\\
\delta\Psi_3&=&-c\bar{\Psi}_2+\delta C_{\mu\nu\lambda\sigma}\bl^\mu \bn^\nu {\bm}^{\!*\,\lambda} \bn^\sigma,\label{dp3}\\
\delta\Psi_4&=&\delta C_{\mu\nu\lambda\sigma}\bn^\mu
{\bm}^{\!*\,\nu} \bn^\lambda {\bm}^{\!*\,\sigma},\label{dp4}\eq
where \bq
a&=&\bm_\mu\delta l^\mu,\label{adef}\\
b&=&\bn_\mu \delta l^\mu + \bl_\mu \delta n^\mu-\bm_\mu\delta {m^*}^\mu - {{\bm}^*}_\mu\delta m^\mu,\label{bdef}\\
c&=&{\bm}^*_\mu \delta n^\mu.\label{cdef}\eq

The gauge invariance of $\delta\Psi_0$ is demonstrated as follows.
(An identical argument applies for $\delta\Psi_4$.) We see from
above that this term depends only on the perturbed Weyl tensor and
on the background tetrad. Both these terms are fixed once the
background and tetrad have been specified and the perturbation has
been added in any particular gauge. Thus $\Psi_0$ is a t.g.i.\
scalar. Then i.g.i.\ follows from the Stewart-Walker lemma
\cite{stewart-walker} (see also Section 1.6 of \cite{stewart})
which we state in this form:
\begin{lemma}
The linearized perturbation of a geometric quantity $Q$ with
background value $\bar{Q}$ is i.g.i.\ iff it satisfies
\[ \mathcal{L}_{\vec{\xi}}\bar{Q} =0 \]
for all generators $\vec{\xi}$ of infinitesmal co-ordinate
transformations of the background space-time.
\end{lemma}
This allows one to characterise all i.g.i.\ quantities
\cite{stewart-walker}:
\begin{lemma}
The linearized perturbation of a geometric quantity $Q$ with
background value $\bar{Q}$ is i.g.i.\ iff one of the following
holds:
\begin{enumerate}
\item $\bar{Q}=0$,
\item $\bar{Q}$ is a constant scalar,
\item $\bar{Q}$ is a constant linear combination of products of
Kronecker deltas.
\end{enumerate}
\end{lemma}

Lemma 1 is trivially satisfied by $\Psi_0$ as it vanishes in the
background; hence full gauge invariance follows. As noted in the
introduction, it is only $\Psi_0$ and $\Psi_4$ which satisfy the
requirements of being both tetrad and identification gauge
invariant. Gauge invariance of these terms has long been
recognized and used; see e.g.\ \cite{chandra}. The form of these
terms in GS variables has been given in \cite{hin} and
\cite{sarbach1}.

Equations (\ref{dp0}-\ref{dp4}) and (\ref{adef}-\ref{cdef})
clearly rule out the possibility of all the Weyl scalars being
t.g.i.\ in general. However if we consider odd and even
perturbations separately, some progress can be made.

\subsection{Odd Perturbations}
In an arbitrary gauge, we have $h_{AB}=h_A^E=G=K=0$ for odd
perturbations. Infinitesmal co-ordinate transformations of odd
parity are generated by 1-form fields of the form (\ref{oddxi}).
We can write down an `odd perturbations only' version of the
Stewart-Walker lemma:
\begin{lemma}
The linearized perturbation of a geometric quantity $Q$ with
background value $\bar{Q}$ is i.g.i.\ with respect to odd
perturbations iff it satisfies
\[ \mathcal{L}_{\vec{\xi}_O}\bar{Q} =0 \]
for all generators $\vec{\xi}_O$ of infinitesmal co-ordinate
transformations of odd parity of the background space-time.
\end{lemma}

The form (\ref{oddxi}) of these generators yields the following
useful result:
\begin{corr}
Let $\bar{S}(x^D)$ and $\bar{T}_{AB\cdots C}(x^D)$ be respectively
a scalar and a covariant tensor field on $M^2$ and define a tensor
field $\bar{T}_{\alpha\beta\cdots\gamma}$ on $M^4$ by padding out
with zeros. Then both $\bar{S}$ and
$\bar{T}_{\alpha\beta\cdots\gamma}$ are i.g.i.\ with respect to
odd perturbations.
\end{corr}
\noindent{\bf Proof} Vanishing of the Lie derivative of $\bar{S}$
along $\vec{\xi}_O$ is immediate. Also, \bs
{\cal{L}}_{\vec{\xi}_O}\bar{T}_{\alpha\beta\cdots\gamma}&=&
\bar{T}_{\alpha\beta\cdots\gamma,\nu}\xi_O^\nu+\bar{T}_{\nu\beta\cdots\gamma}{\xi^\nu_O}_{,\alpha}+\cdots\\
&&+
\bar{T}_{\alpha\beta\cdots\nu}{\xi^\nu_O}_{,\gamma}\nn\\
&=&\bar{T}_{\alpha\beta\cdots\gamma,A}\xi_O^A+\bar{T}_{a\beta\cdots\gamma}{\xi^a_O}_{,\alpha}+\cdots\\&&+
\bar{T}_{\alpha\beta\cdots a}{\xi^a_O}_{,\gamma}\nn\\&=&0.\nn \es

Quantities of particular relevance to us that satisfy this
corollary are the background Weyl scalar ${\bar{\Psi}}_2$ and the
tetrad members $\bar{\bf {l}},\bar{\bf{n}}$. Note that it is
crucial that we consider the tetrad members as 1-forms. Corollary
1 does not apply to contravariant tensor fields.  Hence the
perturbed quantities $\delta l_\mu$, $\delta n_\mu$ are i.g.i.\
with respect to odd perturbations. (Note however that $\delta
l^\mu,\delta n^\mu$ are not i.g.i.) This allows us to make a gauge
invariant choice of the tetrad members $l_\mu,n_\mu$ in the
perturbed space-time. This choice will strongly constrain, in a
gauge invariant manner, the perturbations $\delta m_\mu$ via
(\ref{tetrad}). Furthermore, the parts of $\delta m_\mu$ not fixed
by the choice of $\delta l_\mu$ do not make any contribution to
the perturbed Weyl scalars (\ref{dp0}-\ref{dp4}). Thus subject to
a choice of the i.g.i.\ terms $\delta l_\mu,\delta n_\mu$ (which
is analogous to the choice of tetrad in the background), the
perturbed Weyl scalars are t.g.i.\ When we add in the fact that
$\bar{\Psi}_2$ satisfies Corollary 1, we have our main result.
\begin{prop}
The perturbed Weyl scalars (\ref{dp0}-\ref{dp4}) are
identification and tetrad gauge invariant with respect to odd
perturbations. \end{prop}

We can now calculate these gauge invariant terms. We repeat that
two tetrad choices must be made: (i) we specify that the
background tetrad uses the principal null directions as its real
members and (ii) we must specify the gauge invariant terms $\delta
l_\mu,\delta n_\mu$. We note however that $\delta\Psi_0$ and
$\delta\Psi_4$ depend only on the first choice. In fact the same
is true for $\delta\Psi_2$: using (\ref{tetrad}), we can show that
\[ b=-\bg^{\mu\nu}\delta g_{\mu\nu}.\] Thus there is no
contribution to $\delta\Psi_2$ from the perturbed tetrad.

The most obvious gauge invariant choice for the perturbation of
the real members of the null tetrad is $\delta l_\mu=\delta
n_\mu=0$. Working in the RW gauge, we can then solve
(\ref{tetrad}) for $\delta m_\mu$; as noted above, {\em any}
particular solution of this system yields the same Weyl scalars.
Then we calculate the Weyl scalars, and to conclude, write these
in terms of the i.g.i.\ quantities of Section II. The result is
 \bq
\delta\Psi_0&=&\left.\frac{Q_0}{2r^2}\right.\bl^A\bl^Bk_{A|B},\label{oddp0}\\
\delta\Psi_1&=&\left.\frac{Q_1}{r}\right.\left[(r^2\Pi)_{|A}\bl^A-\frac{4}{r^2}k_A\bl^A\right],\label{oddp1}\\
\delta\Psi_2&=&{Q_2}\Pi,\label{oddp2}\\
\delta\Psi_3&=&\left.\frac{Q^*_1}{r}\right.\left[(r^2\Pi)_{|A}\bn^A-\frac{4}{r^2}k_A\bn^A\right],\label{oddp3}\\
\delta\Psi_4&=&\left.\frac{Q^*_0}{2r^2}\right.\bn^A\bn^Bk_{A|B},\label{oddp4}
\eq where
\[ \Pi = \epsilon^A_B(r^{-2}k^B)_{|A}\]
is the scalar introduced in \cite{GS} which appears in the master
equations for odd perturbations. The angular coefficients here are
given by
\begin{eqnarray}
Q_0&=&-2w^aw^bS_{a:b},\\
Q_1&=&-\frac14 w^aS_a,\\
Q_2&=&-\frac{i}{4}l(l+1)Y,\\
\end{eqnarray}
where $w^a=r^{-1}\bar{m}^a.$   We can now give an interpretation
of the gauge invariant metric perturbation $k_A$ based on these
scalars using the work of Szekeres \cite{szek}. The scalars
$\Psi_0,\Psi_4$ are independent of the choice of perturbation in
the tetrad and so depend only on our choice of background tetrad
which, as argued above, may be considered to be invariant. Thus
these two terms represent pure transverse gravitational waves
propagating in the radial inward (respectively outward) null
directions. We note that the formulae (\ref{oddp0}) and
(\ref{oddp4}) have been given previously in \cite{sarbach1}.

Similarly, $\Psi_2$ is independent of the choice of tetrad
perturbation. Thus this term invariantly describes a perturbation
of the Coulomb component of the gravitational field.

The scalars $\Psi_1,\Psi_3$ depend on the choice of tetrad
perturbation. However with our gauge invariant choice described
above, we can state that the relevant coefficients represent pure
longitudinal gravitational waves propagating in the radial inward
(respectively outward) null directions.

We note that these statements are valid for $l\geq 2$. The angular
coefficient $Q_0$ vanishes identically for $l=1$. Thus the
vanishing of the terms $\delta\Psi_0$ and $\delta\Psi_4$ for $l=1$
is gauge invariant (and of course entirely expected: we only
expect these gravitational radiation terms to switch on for the
quadrupole and higher moments, $l\geq 2$). For $l=1$, $\Pi$ is
i.g.i\, but $k_A$ is not so. Hence $\delta\Psi_2$ is gauge
invariant, but $\delta\Psi_1$ and $\delta\Psi_3$ are not.

We note also that (\ref{oddp0})-(\ref{oddp4}) completely specify
the gauge invariant metric perturbation; that is, these equations
may be solved for $k_A$ in terms of $\delta\Psi_{1-4}$. In
particular, vanishing of the perturbed Weyl scalars at a point of
space-time implies vanishing of $k_A$ at that point.

\subsection{Even perturbations}
For even perturbations, we set $h_A^O=h=0$. Infinitesmal
co-ordinate transformations of even parity are generated by
1-forms of the form (\ref{evenxi}). The `even perturbations only'
version of Lemma 3 is immediate. The following result describes
the terms additional to those described by Lemma 2 which become
i.g.i.\ when we restrict to even perturbations.
\begin{lemma}
Let $\bar{Q}(x^\mu)$ and $\bar{v}_\mu(x^\nu)$ be respectively a
scalar and a 1-form defined on $M^4$. Then the linear
perturbations of $\bar{Q}$ and $\bar{v}_\mu$ are i.g.i.\ with
respect to even perturbations iff $\bar{Q}=\bar{Q}(x^a)$ with \[
\gamma^{ab}\bar{Q}_{,a}Y_{b}=0,\] $\bar{v}_A=0$ and
\[ \bar{v}_a=\lambda S_a, \]
where $\lambda(x^b)$ satisfies
\[ Y_aY^a\lambda_{,b}Y^b+Y_{a:b}(Y^aY^b-S^aS^b)\lambda=0. \]
There are no vector fields $\bar{v}^\mu$ which are i.g.i.\ with
respect to even perturbations.
\end{lemma}
Note that it possible to construct covariant tensor fields of
higher rank which are i.g.i.\ by taking tensor products of the
1-forms described by the lemma.

\noindent{\bf Proof:} The proof for the scalar case is immediate.
In the 1-form case, the result follows by writing down the
equations ${\cal{L}}_{\vec{\xi}_E} \bar{v}_\mu=0$. This equation
must hold for all $\vec{\xi}_E$ with 1-form equivalents given by
(\ref{evenxi}). We obtain $\bar{v}_A=0$ by considering particular
forms of $\xi^\mu$. We also obtain $\bar{v}_a=\bar{v}_a(x^b)$ and
$Y^a\bar{v}_a=0$. Since we are in 2 dimensions and $Y^aS_a=0$,
this implies that we can write $\bar{v}_a=\lambda(x^b)S_a$. The
remaining conditions reduce to the linear p.d.e.\ for $\lambda$
given in the statement.

Unlike the corresponding situation for odd perturbations, there is
no hope of constructing useful gauge invariant background terms
from the quantities described in this lemma. In particular, it is
not possible to use the 1-forms described in the lemma to
construct some of the null tetrad members. This is essentially
because one cannot have any $x^A$ dependence in the gauge
invariant terms. Thus we can summarise as follows.

\begin{prop}
$\delta\Psi_0$ and $\delta\Psi_4$ are the only perturbed Weyl
scalars that are identification and tetrad gauge invariant with
respect to even perturbations.
\end{prop}
 For completeness, we give these terms which have been given
 previously in \cite{sarbach1}:
\bq \delta\Psi_0 &=& \frac{1}{2r^2}\bar{l}^A\bar{l}^B k_{AB}
(w^aw^bY_{:ab}),\\
\delta\Psi_4 &=&\frac{1}{2r^2}\bar{n}^A\bar{n}^B k_{AB}
(w^{\!*\,a}w^{\!*\,b}Y_{:ab}). \eq

For the lowest multipole moments $l=0,1$, the angular coefficients
here vanish identically, and so the vanishing of $\delta\Psi_0$
and $\delta\Psi_4$ is gauge invariant.

\section{Conclusions}
We have investigated the possibility of giving a gauge invariant
physical interpretation of gauge invariant metric perturbations of
spherically symmetric space-times by considering the perturbed
Weyl scalars. This turns out to be possible only for the case of
odd perturbations; however in this case, it transpires that all
the perturbed Weyl scalars are identification and tetrad gauge
invariant, and so the physical interpretation of the metric terms
can be made. One can therefore immediately see the contribution of
a particular metric perturbation to ingoing and outgoing
longitudinal and transverse gravitational waves, and to the
Coulombic interaction term. We anticipate that this will be of use
in various different studies, for example in our ongoing work on
the stability of Cauchy horizons in self-similar collapse
\cite{bn+tw}. The expressions (\ref{oddp0}-\ref{oddp4}) can be
used to set co-ordinate independent and gauge invariant boundary
conditions for perturbations, and can also be used as indicators
of instability in different regimes (for example if such terms
diverge in the approach to a singularity or to a Cauchy horizon).
Care is needed here however. While the terms
(\ref{oddp0}-\ref{oddp4}) indicate the presence or otherwise of
various gravitational waves and Coulomb-type perturbations, they
should not be used to determine magnitudes. This is crucial in
setting boundary conditions, where one typically imposes a
condition on the limiting behaviour of a physically significant
quantity. This is because of the scale co-variance in the scalars
resulting from the spin-boosts (\ref{boosts}): under these Lorentz
transformations, we have
\[ \delta\Psi_n\to a^{2-n}\delta\Psi_n,\quad n=0,\dots,4.\]
(For convenience, we have set $\omega=0$ in (\ref{boosts}) as this
will not affect magnitudes.) However this shows that the following
g.i.\ first-order quantities have physically significant
magnitudes, and so can be used for setting boundary conditions:
\bs \delta P_{-1} &=& |\delta\Psi_0\delta\Psi_4|^{1/2},\\ \delta
P_0 &=& \delta\Psi_2,\\ \delta
P_1&=&|\delta\Psi_1\delta\Psi_3|^{1/2}. \es All three provide
terms useful for the analysis of odd perturbations, while the
first can also be used for even perturbations (and indeed in more
general contexts \cite{beetle}).

\section*{Acknowledgement}
I thank Thomas Waters for useful conversations. This research is
supported by Enterprise Ireland grant SC/2001/199.

\end{document}